\documentclass[10pt,conference]{IEEEtran}

\bstctlcite{IEEEexample:BSTcontrol}

\usepackage{epsfig,rotating,setspace,latexsym,amsmath,epsf,amssymb,amsfonts,bm,theorem,cite,authblk, bbm,color, multirow, caption, subcaption, booktabs, xcolor}
\IEEEoverridecommandlockouts
\allowdisplaybreaks

\title{When Semantic Communication Meets Queueing: Cross-Layer Latency and Task Fidelity Optimization}

\author{Yalin E. Sagduyu and Tugba Erpek}
\affil{\normalsize Nexcepta, Gaithersburg, MD, USA}

\date{}
\begin{document}
\bstctlcite{BSTcontrol}

\maketitle
\thispagestyle{empty}

\begin{abstract}
Semantic communication (SemCom) with learned encoder-decoder architectures enables end-to-end learning of compact task-oriented representations optimized for the wireless channel, reducing channel resources needed to convey task-relevant information and improving spectrum efficiency. This paper studies semantic image transmission over block Rayleigh fading with AWGN using a multi-task semantic autoencoder that jointly reconstructs images and predicts labels from the received waveform. The latent dimension (complex channel uses per source sample) serves as a cross-layer control variable governing semantic fidelity and channel resource usage. We characterize the resulting latency-task fidelity tradeoff: larger latent representations improve inference accuracy but increase service time, channel uses, and queueing delay. Building on this insight, we develop online semantic-rate controllers that adapt the latent dimension per update under a long-term semantic error constraint. A queue-aware drift-plus-penalty policy minimizes delay subject to an average semantic error cap, while a complementary age-aware policy minimizes time-average Age of Information (AoI). By adapting the semantic rate to congestion and fidelity requirements, the proposed framework improves spectrum utilization and enables timely semantic updates with significantly lower delay and AoI than fixed-rate baselines.
\end{abstract}

\begin{IEEEkeywords}
Semantic communications, deep neural networks, latency, spectrum efficiency, cross-layer design.
\end{IEEEkeywords}

\section{Introduction}
Wireless spectrum is scarce, particularly for emerging applications such as autonomous systems, edge intelligence, and distributed sensing. Conventional communication systems reliably deliver all source bits, even when many are irrelevant to the end task, leading to inefficient spectrum utilization when the objective is inference, detection, or situational awareness rather than exact signal reconstruction. Semantic communication (SemCom) addresses this inefficiency by transmitting compact task-relevant representations instead of raw data, reducing channel uses per update and improving spectrum efficiency and access to shared wireless resources.

SemCom optimizes transmission for task-relevant meaning, rather than exact symbol recovery \cite{Gunduz2023BeyondBits,Uysal2022SemanticNetworks,Sagduyu2024Will6G}. For high-dimensional sources such as images, learned encoder-decoder models map each sample to a compact latent representation transmitted over the wireless channel and decoded into semantic outputs \cite{Bourtsoulatze2019DeepJSCC,Jankowski2021WirelessRetrieval,Erdemir2023GenerativeJSCC,Lyu2024SemanticImage}. The latent dimension (complex channel uses per sample) defines the semantic rate: increasing it improves task fidelity and robustness but also increases transmission time and spectrum consumption, raising latency and limiting the number of semantic updates. This fidelity-service-time coupling is critical in time-sensitive systems such as autonomous driving, edge video analytics, and UAV situational awareness, motivating joint optimization of communication, computation, and learning resources and highlighting the latent dimension as a key cross-layer control variable.

From a network perspective, the latent dimension determines the spectrum resources consumed by each semantic update. Large latent representations waste channel uses and limit supported updates, while aggressive compression degrades semantic fidelity, making adaptive semantic-rate control essential for efficient spectrum utilization. Motivated by latency-critical sensing and situational awareness, we study semantic image transmission over block Rayleigh fading with AWGN. We develop an end-to-end semantic autoencoder that preserves semantic meaning at the receiver and characterize classification accuracy versus signal-to-noise ratio (SNR) and latent dimension, revealing saturation where additional channel uses yield diminishing gains but higher latency. Prior work highlights freshness and representation efficiency \cite{Sagduyu2023AoI,Sagduyu2024MultiRound,singh2024computingtradeoff,Yang2023EnergyEfficientSemantic,sagduyu2024joint}, but largely assumes fixed semantic rates, leaving online control under stochastic traffic and fidelity constraints underexplored.

The central contribution of this work is a cross-layer control framework that treats the latent dimension as the action for each semantic update and adapts the semantic rate online under a long-term semantic fidelity requirement. We impose an average semantic error constraint and develop Lyapunov drift-plus-penalty controllers: a queue-aware policy that minimizes delay and a freshness-aware policy that minimizes time-average Age of Information (AoI). Both operate without retransmission and enforce the error cap using a virtual fidelity queue. Compared with prior SemCom designs that optimize distortion or task accuracy, the proposed framework exposes the accuracy-latency-freshness tradeoff induced by the latent dimension and enables near-optimal online operation under stochastic arrivals. Fixed-latent systems are inefficient: small latent dimensions underutilize semantic fidelity under favorable channels, while large latent dimensions incur unnecessary delay when fidelity pressure is low. The proposed cross-layer adaptation resolves this inefficiency by dynamically matching the semantic rate to queue state and fidelity debt.

The remainder of the paper is organized as follows. Sec.~\ref{sec:queue_sem_rate} describes the SemCom system. Sec.~\ref{sec:delay} presents queue-aware semantic rate control under fidelity constraints. Sec.~\ref{sec:aoi} extends the formulation to AoI. Sec.~\ref{sec:conclusion} concludes the paper.

\section{Semantic Communications} \label{sec:queue_sem_rate}
\begin{figure}[t!]
  \centering
 \includegraphics[width=\columnwidth]{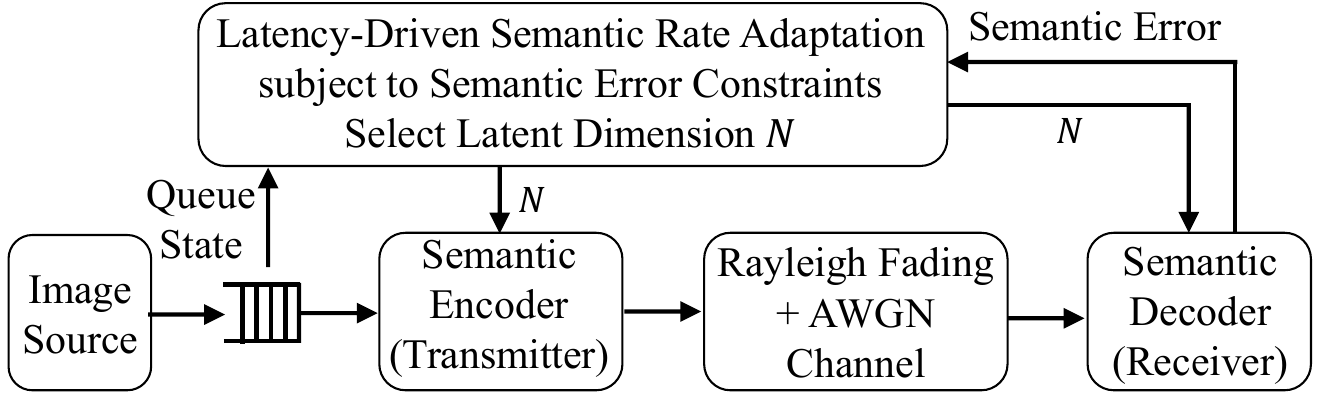}
 \vspace{0.05cm}
  \caption{SemCom system model.}
  \label{fig:system_model} 
\end{figure}

The system model of SemCom, shown in Fig.~\ref{fig:system_model}, consists of a semantic encoder-decoder communicating over the wireless channel, with rate adaptation introduced in Secs.~\ref{sec:delay}-\ref{sec:aoi}.

\subsection{Semantic Encoder}
Let $\mathbf{x}$ denote an input image with label $l \in\{0,\ldots,L-1\}$, and let $\tilde{\mathbf{x}}$ denote its normalized version with pixel values in $[0,1]$. The semantic encoder $f_{\mathrm{enc}}$ maps $\tilde{\mathbf{x}}$ to a complex baseband sequence
\begin{equation}
\mathbf{s}=f_{\mathrm{enc}}(\tilde{\mathbf{x}})\in\mathbb{C}^{N},
\end{equation}
where $N$ denotes the number of complex channel uses per image. Smaller $N$ enforces stronger compression and semantic abstraction, while larger $N$ provides additional degrees of freedom, improving semantic fidelity and robustness to fading. The encoder is implemented as a deep convolutional neural network (CNN) with residual learning and channel-wise attention. An initial convolutional stem extracts low-level spatial features. This is followed by multiple stages of residual blocks that progressively increase channel depth while reducing spatial resolution through strided convolutions. Each residual block consists of two convolutional layers with batch normalization and ReLU activation, together with an identity shortcut that stabilizes gradient flow in deep networks.

Within each residual unit, squeeze-and-excitation (SE) gating uses spatial pooling and a two-layer fully connected network with a reduction-and-expansion structure to generate adaptive channel weights, enabling the encoder to prioritize semantically informative features, especially when $N$ is small. After the final convolutional stage, global average pooling produces a compact embedding vector $\mathbf{e}$. Two fully connected layers map this embedding to a real-valued vector $\mathbf{u}\in\mathbb{R}^{2N}$, which is reshaped into a complex baseband sequence $\mathbf{s}\in\mathbb{C}^N$ with elements
\begin{equation}
s_k = u_{2k-1} + j\,u_{2k},\qquad k=1,\ldots,N.
\end{equation}
The resulting sequence is normalized per sample to satisfy a unit average power constraint, ensuring a fixed transmit SNR across latent dimensions.

\subsection{Channel Model and Receiver-Side Conditioning}
Transmission occurs over a block Rayleigh fading channel
\begin{equation}
\mathbf{y}=h\,\mathbf{s}+\mathbf{n},\qquad
h\sim\mathcal{CN}(0,1),\quad
\mathbf{n}\sim\mathcal{CN}(\mathbf{0},\sigma^2\mathbf{I}_N),
\end{equation}
where a single fading coefficient $h$ applies across the $N$ channel uses of each image, $\mathbf{n}$ is additive complex Gaussian noise with variance $\sigma^2$ set by the target SNR under unit average transmit power. The receiver produces both a reconstructed image $\hat{\mathbf{x}}$ and semantic logits $\hat{\mathbf{z}}\in\mathbb{R}^L$ via learned decoders. To mitigate fading-induced variation under noncoherent operation, the received latent waveform $\mathbf{y}$ is normalized per sample, and a conditioning vector $\mathbf{c}$ is derived from its low-order statistics. No receiver channel state information is assumed; robustness follows from normalization and conditioning.

\subsection{Reconstruction Decoder}

The reconstruction decoder maps the received latent representation to an image estimate with the same spatial dimensions as the input (e.g., 32$\times$32$\times$3 for CIFAR-10). The normalized received latent sequence $\mathbf{y}$ is flattened, concatenated with a receiver-side conditioning vector $\mathbf{c}$, and passed through two fully connected layers to produce an initial spatial feature map. This map is progressively upsampled using interpolation followed by residual convolutional refinement blocks with channel attention, enabling recovery of spatial structure while avoiding checkerboard artifacts. A final convolution with a bounded activation produces valid pixel values. When the latent dimension $N$ is small, reconstruction is underdetermined and relies more on learned image priors, preserving global semantics but losing high-frequency detail. Increasing $N$ improves Peak Signal-to-Noise Ratio (PSNR) and Structural Similarity Index Measure (SSIM), reducing reconstruction ambiguity.

\subsection{Semantic Classifier Decoder}

In parallel, a semantic classifier infers labels from the received waveform. The receiver forms a multi-channel representation from the normalized received latent sequence $\mathbf{y}$. A stack of 1D convolutions processes the sequence along the symbol dimension to extract waveform features. The resulting features are globally pooled, reweighted via channel-wise gating, and augmented with the conditioning vector $\mathbf{c}$ to improve robustness to fading. The features are then passed through a fully connected network to produce class logits. Larger $N$ increases the effective temporal context, improving robustness at the cost of more channel uses. The predicted label is given by $\hat{l} = \arg\max_{i} \hat{z}_i$.

\subsection{End-to-End Training}
The system is trained using a weighted composite loss that balances reconstruction fidelity and semantic accuracy. The loss consists of a reconstruction term combining mean-squared error (MSE) with a small $\ell_1$ component to capture global fidelity and edge sharpness, a structural term based on SSIM to preserve local contrast and texture, a classification term based on cross-entropy, and an auxiliary PSNR-based term that complements MSE by capturing reconstruction quality on a logarithmic scale. Training uses stochastic gradient optimization. For each mini-batch, images are encoded into $N$ complex symbols, transmitted over the channel, and processed by both decoders. The composite loss is backpropagated through the channel to update all parameters. Training uses the Adam optimizer with batch size 128 and is implemented on NVIDIA RTX PRO 6000 Blackwell GPUs.

\begin{figure}[t!]
  \centering
 \includegraphics[width=0.8\columnwidth]{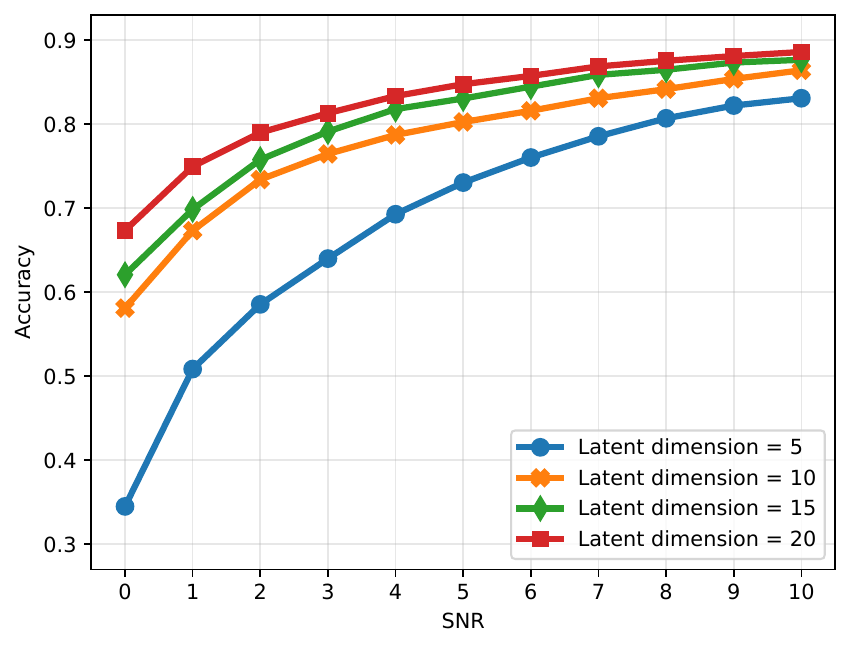}
 \vspace{-0.1cm}
  \caption{Accuracy vs.\ SNR in SemCom.}
  \label{fig:acc_vs_snr} 
  \vspace{0.25cm}
 \includegraphics[width=0.8\columnwidth]{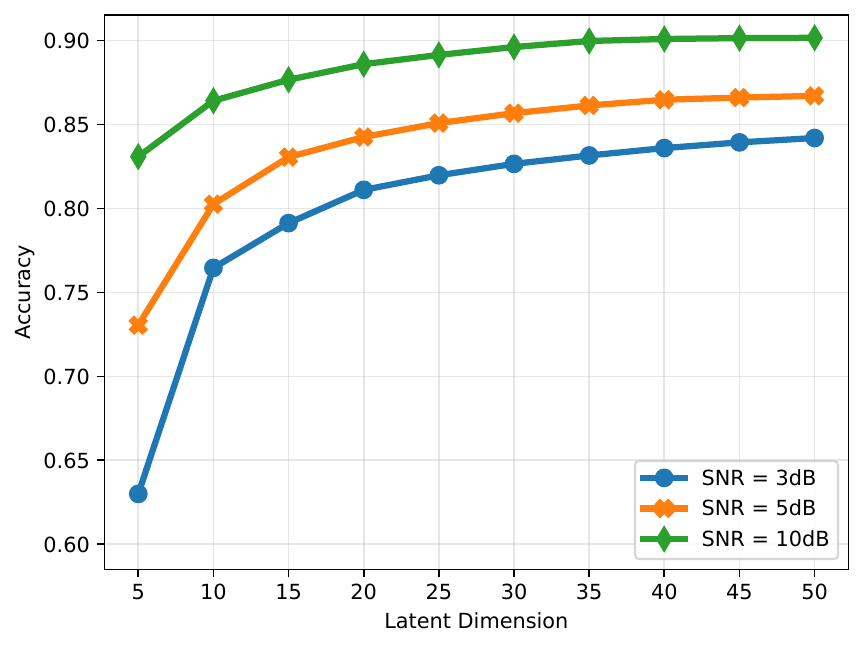}
 \vspace{-0.1cm}
  \caption{Accuracy vs.\ latent dimension in SemCom.}
  \label{fig:acc_vs_latentdim}
\end{figure}

\subsection{SemCom Performance and Latent-Dimension Tradeoff}
We use the CIFAR-10 dataset with $L=10$ classes for training and evaluation. We first examine classification accuracy versus SNR for several latent dimensions. As shown in Fig.~\ref{fig:acc_vs_snr}, accuracy increases with SNR for all $N$ due to reduced channel distortion, with larger latent dimensions consistently achieving higher accuracy. The gap is largest at low SNR, reflecting implicit redundancy benefits, while saturation occurs at high SNR once semantic information is preserved. 

Fig.~\ref{fig:acc_vs_latentdim} shows accuracy versus latent dimension for several SNRs. Accuracy increases with $N$ and then saturates, indicating diminishing gains per additional channel use. Larger $N$ improves decoding fidelity but increases service time approximately linearly, leading to higher delay. Thus, small $N$ yields low delay but reduced fidelity, whereas large $N$ improves accuracy at higher delay. Fixed-latent designs are therefore inefficient: large $N$ wastes resources under favorable conditions, while small $N$ fails under deep fades or strict fidelity targets, suggesting operation near the saturation region rather than at maximum latent dimension. Reconstruction remains effective with PSNR 18.16\,dB and SSIM 0.60. Overall, the latent dimension $N$ acts as the key control variable for semantic fidelity, improving accuracy at the cost of latency and motivating adaptive cross-layer semantic rate tuning.

\section{Queue-Aware Semantic Rate Control} \label{sec:delay}

This section develops a queue-aware SemCom controller that treats the latent dimension $N$ as the per-update action. It minimizes delay under an average semantic error constraint via Lyapunov drift-plus-penalty control.

\subsection{Queueing Model and Action Space}

Consider a single-server queue with exogenous Poisson arrivals $\{A(t)\}$ at rate $\lambda$ images per unit time, where $A(t)$ denotes the number of arrivals in $[0,t]$. Let $Q(t)$ denote the number of waiting images at time $t$ and $Q_{\mathrm{sys}}(t)$ denote the total number in the system (including the one in service):
\begin{equation}
Q_{\mathrm{sys}}(t) = Q(t) + \mathbf{1}\{\text{server busy}\}.
\end{equation}
We use $Q_{\mathrm{sys}}(t)$ for all control decisions and delay analysis.

Let $\{t_k\}_{k\ge1}$ denote the service start times, and let $k$ index the image entering service at time $t_k$. At each service start time $t_k$, the controller selects a latent dimension
\begin{equation}
N_k \in \mathcal{N} \triangleq \{N_1,\ldots,N_M\},
\end{equation}
where $N_k$ is the number of complex channel uses allocated to the $k$th image and thereby determines its service time. Transmitting the $k$th image with latent dimension $N_k$ consumes $N_k$ channel uses and therefore incurs deterministic service time
\begin{equation}
S_k = N_k.
\end{equation}
This deterministic service-time model follows from time normalization and isolates the impact of the latent dimension on delay. The control process $N(t)$ is piecewise constant, taking value $N_k$ during service of the $k$th image. For each $N\in\mathcal{N}$ at a fixed operating SNR under Rayleigh fading, the semantic error probability is denoted by $p_e(N)\in[0,1]$, averaged over channel realizations and data samples, and typically decreasing with $N$. Realized semantic errors for the $k$th served image are modeled as Bernoulli indicators $E_k\in\{0,1\}$ with
\begin{equation}
\mathbb{P}(E_k=1 \mid N_k=N) = p_e(N).
\end{equation}
Errors are conditionally independent across transmissions given the selected latent dimension. No retransmission is employed; each image is transmitted once without feedback.

\subsection{Minimum Delay Subject to Average Error Constraint}

We seek an online policy that stabilizes the queue while enforcing a long-term semantic fidelity constraint. Let $E_k\in\{0,1\}$ denote the semantic error indicator of the $k$th served image, and define the long-run average semantic error rate
\begin{equation}
\bar{E} \triangleq \limsup_{K\to\infty}\frac{1}{K}\sum_{k=1}^{K}\mathbb{E}[E_k].
\end{equation}
The semantic fidelity requirement is
\begin{equation}
\bar{E} \le \varepsilon,
\end{equation}
for a prescribed semantic error cap $\varepsilon\in(0,1)$.

The latency objective is to minimize average delay. By Little's law, under stability the long-term average delay satisfies
\begin{equation}
\bar{W} = \frac{\bar{Q}_{\mathrm{sys}}}{\lambda},
\end{equation}
so minimizing delay is equivalent to minimizing the average system backlog $\bar{Q}_{\mathrm{sys}}$. Combining these, we consider the constrained stochastic optimization problem:
\begin{equation}
\min \:\: \bar{Q}_{\mathrm{sys}} \:\:\:\:
\text{s.t.} \:\:\:\: \bar{E} \le \varepsilon, \:\: N_k\in\mathcal{N}, \: \:  \sup_{t}\mathbb{E}[Q_{\mathrm{sys}}(t)] < \infty,
\end{equation}
where queue stability means that the expected queue length remains bounded. This formulation explicitly captures the semantic fidelity-latency tension. Selecting larger $N$ reduces $p_e(N)$ but increases the service time and thus queueing delay.

\subsection{Virtual Queue for the Semantic Fidelity Constraint}
To enforce $\bar{E}\le\varepsilon$ online, we introduce a virtual queue $Z(k)\ge 0$ updated per served image using the realized indicator $E_k$. Define
\begin{equation}
Z(k{+}1) = \max\left\{ Z(k) + E_k - \varepsilon,\ 0 \right\}.
\label{eq:virtualZ}
\end{equation}
This is a standard deficit queue. If $E_k=1$, the virtual backlog increases by $1-\varepsilon$; otherwise it decreases by $\varepsilon$ (clipped at $0$). Under mild ergodicity conditions, mean-rate stability of $Z(k)$ ensures the time-average constraint:
\begin{equation}
\limsup_{K\to\infty}\frac{1}{K}\sum_{k=1}^{K}\mathbb{E}[E_k] \le \varepsilon.
\end{equation}
$Z(k)$ acts as an adaptive Lagrange multiplier, rising for aggressive (small $N$) actions and falling for conservative ones.

\subsection{Lyapunov Drift-Plus-Penalty Control Law}

We derive a low-complexity online policy using Lyapunov optimization. Let the Lyapunov function be
\begin{equation}
L(Q_{\mathrm{sys}},Z)=\tfrac12 Q_{\mathrm{sys}}^2 + \tfrac12 Z^2.
\end{equation}
At each service decision epoch, we select $N$ to approximately minimize an upper bound on the one-step conditional drift plus a service-time penalty. Intuitively, this reduces both backlogs $Q_{\mathrm{sys}}$ and $Z$ while penalizing large $N$ that increases delay.

A standard drift analysis yields a myopic rule of the form
\begin{equation}
N_k \in \arg\min_{N\in\mathcal{N}}
\left\{ Q_k \, N + V\, Z_k \, \hat p_e(N) \right\},
\label{eq:policy}
\end{equation}
where $\hat p_e(N)$ is an estimate of the semantic error probability $\mathbb{P}(E_k=1\!\mid\! N_k=N)$, $Q_k \triangleq Q_{\mathrm{sys}}(t_k)$ is the number of images in the system at the $k$th service-start epoch (including the image in service), $Z_k$ is the virtual backlog, and $V\!\ge\!0$ is a tunable weight trading delay against constraint enforcement. The term $Q_k N$ penalizes long service times under congestion, favoring smaller $N$, while $Z_k \hat p_e(N)$ penalizes unreliable actions when the virtual queue is large, favoring larger $N$ to reduce expected semantic error. Increasing $V$ strengthens fidelity enforcement, typically lowering average error at the cost of higher delay.

\subsection{Event-Driven Queue Dynamics and Delay Estimation}
Service is event-driven with deterministic service time $N_k$. Arrivals follow the exogenous process and departures occur at completion. Let $Q_{\mathrm{sys}}(t)$ denote the number of images in the system (waiting plus the one in service). The time-average queue length is computed via the renewal-reward identity:
\begin{equation}
\bar{Q}_{\mathrm{sys}} \approx \frac{1}{T}\int_{0}^{T} Q_{\mathrm{sys}}(t)\,dt,
\end{equation}
which is obtained by accumulating the area under $Q_{\mathrm{sys}}(t)$ between events. The average delay follows from Little’s law,
\begin{equation}
\bar{W} \approx \frac{\bar{Q}_{\mathrm{sys}}}{\lambda},
\end{equation}
provided the system is stable and the horizon is long. 

\subsection{Minimum Delay Subject to Semantic Fidelity Constraint}

The control law \eqref{eq:policy} yields a family of policies indexed by $V$. For fixed arrival rate $\lambda$ and error cap $\varepsilon$, one can evaluate the achieved pair $(\bar{W}(V), \bar{E}(V))$ and identify the minimum delay among those satisfying the semantic fidelity requirement:
\begin{equation}
\bar{W}^\star(\lambda,\varepsilon) = \min_{V\ge 0}\ \bar{W}(V)
\quad \text{s.t.}\quad \bar{E}(V) \le \varepsilon.
\end{equation}
This traces the accuracy-latency frontier. Small $V$ favors low latency but may violate the error constraint, while large $V$ prioritizes semantic fidelity at higher delay. The controller adapts the semantic rate based on the estimated semantic error, using larger $N$ only when needed to meet the fidelity constraint and smaller $N$ otherwise to reduce delay.

The mechanism is a coupled primal-dual system linking SemCom and queueing dynamics. SemCom provides $\mathcal{N}$ and $p_e(N)$, the queue provides the state $Q_k$, and the virtual queue $Z_k$ enforces the error constraint. The lightweight policy \eqref{eq:policy} adapts in real time to congestion and fidelity debt. Since $N$ jointly controls latency and semantic fidelity, drift-plus-penalty control minimizes delay under the error bound.

\subsection{Delay Performance Under Semantic Fidelity Constraints}
\begin{figure}[t!]
  \centering
 \includegraphics[width=0.8\columnwidth]{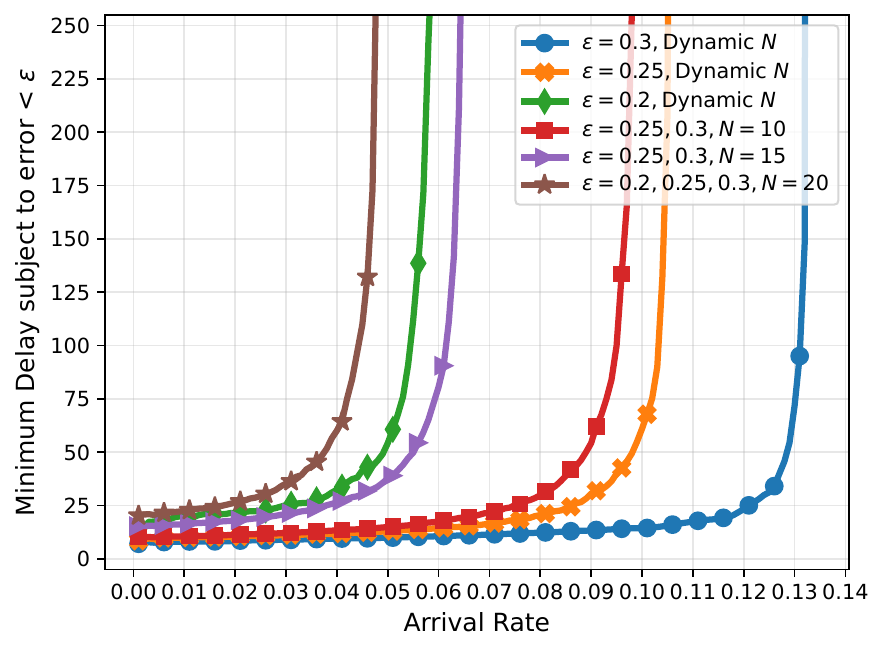}
  \caption{Delay vs. arrival rate.}
  \label{fig:delay_vs_arrival}
  \vspace{-0.4cm}
\end{figure}

We evaluate the queue-aware semantic rate controller under varying traffic loads and fidelity requirements. Fig.~\ref{fig:delay_vs_arrival} shows average delay versus arrival rate $\lambda$ for the adaptive latent-dimension policy and fixed-$N$ baselines under different error caps $\varepsilon$. The results highlight the fidelity-latency coupling. Delay increases slowly at light load and rises sharply as $\lambda$ approaches service capacity, consistent with single-server behavior where delay scales as $(1-\rho)^{-1}$ with utilization $\rho=\lambda \mathbb{E}[N]$, where $\mathbb{E}[N]$ is the average latent dimension selected by the policy. The sharp upturn indicates the onset of instability as service time becomes insufficient for the arrival load. Tightening the fidelity cap from $\varepsilon=0.3$ to $0.25$ and $0.2$ shifts delay curves upward and shrinks the stability region, since stricter fidelity requires more frequent use of larger $N$, increasing mean service time. The figure therefore quantifies the cost of enforcing semantic fidelity in SemCom.

The dynamic policy achieves lower delay than fixed-$N$ baselines by adapting $N$ using the backlog and virtual fidelity queue, selecting smaller $N$ when backlog is low and increasing $N$ only to maintain $\bar{E}\le\varepsilon$. Fixed-$N$ schemes show clear feasibility limits: $N=10$ supports only $\varepsilon=0.3$, $N=15$ meets $\varepsilon=0.25$ but struggles at $\varepsilon=0.2$, and only $N=20$ satisfies $\varepsilon\in\{0.2,0.25,0.3\}$ at much higher delay. Thus, no fixed semantic rate achieves both low delay and strict fidelity. Near the stability boundary, fixed-$N$ schemes diverge earlier, whereas the adaptive policy compresses more aggressively under backlog, extending the feasible arrival region while maintaining the constraint. The results confirm that the latent dimension $N$ is the key cross-layer control variable, with adaptive control approaching the optimal fidelity-latency balance, while fixed-$N$ designs either violate fidelity or increase delay.

\section{AoI-Aware Semantic Rate Control} \label{sec:aoi}

We develop an AoI-aware controller that adapts the semantic rate (latent dimension) to minimize time-average AoI under a semantic fidelity constraint, directly targeting freshness.

\subsection{AoI System Model}
Consider the same single-server SemCom system with instantaneous AoI $\Delta(t)$ defined as the time since the most recently successfully delivered image was generated. The queue operates in FIFO order, with each update represented by its generation time. Let $g_k$ and $d_k$ denote the generation and departure times of the $k$th update, and $E_k\in\{0,1\}$ the semantic error indicator ($1$ failure, $0$ success). The AoI resets only upon successful decoding:
\begin{equation}
\Delta(d_k^+) =
\begin{cases}
d_k - g_k, & E_k=0,\\
\Delta(d_k^-), & E_k=1,
\end{cases}
\end{equation}
where $\Delta(d_k^-)$ denotes the AoI just before completion of the $k$th service. Between events, AoI increases linearly with unit slope, i.e., $\frac{d}{dt}\Delta(t)=1$. The objective is the time-average AoI
\begin{equation}
\bar{\Delta} \triangleq \limsup_{T\to\infty} \frac{1}{T}\int_0^T \Delta(t)\,dt.
\end{equation}

As before, selecting latent dimension $N\in\mathcal{N}$ determines both the service time and the semantic error probability $p_e(N)$ in the same cross-layer coupling. Larger $N$ improves semantic fidelity but increases service time and may worsen freshness.

\subsection{Optimization Problem and Virtual Semantic Fidelity Queue}

The AoI-aware semantic control problem is
\begin{equation}
\min \: \bar{\Delta} \quad 
\text{s.t.}\quad  \bar{E} \le \varepsilon, \:\: N_k\in\mathcal{N}, \: \: \{Q_{\mathrm{sys}}(t)\}\ \text{is stable}.
\end{equation}

Compared with the delay formulation in Section~\ref{sec:queue_sem_rate}, the key difference is that the controller must account for the age penalty incurred by long service times. We reuse the same virtual queue to enforce the semantic error constraint. At each service completion $k$,
\begin{equation}
Z(k{+}1) = \max\!\left\{ Z(k) + E_k - \varepsilon,\ 0 \right\}.
\end{equation}
As discussed earlier, stability of $Z(k)$ implies satisfaction of the long-term semantic fidelity requirement. The key distinction from the delay-focused design lies in the per-decision cost used in the drift-plus-penalty step.

\subsection{AoI-Driven Drift-Plus-Penalty Policy}
Using standard AoI sample-path arguments, the incremental AoI cost incurred by serving a semantic update of duration $N$ when the current age is $\Delta$ is
\begin{equation}
\Delta N + \frac{1}{2}N^2,
\end{equation}
which captures the trapezoidal area accumulated during service. Following the same Lyapunov optimization principle used earlier, but replacing the backlog penalty with the AoI cost, the online semantic-rate decision becomes
\begin{equation}
N_k \in \arg\min_{N\in\mathcal{N}}
\left\{
\Delta_k N + \frac{1}{2}N^2
+ V\, Z_k\, \hat p_e(N)
\right\}.
\label{eq:aoi_policy}
\end{equation}
Here, $\Delta_k \triangleq \Delta(t_k^-)$ denotes the AoI at the start of the $k$th service. This rule differs from the queue-driven policy in Section~\ref{sec:queue_sem_rate}. The term $\Delta_k N$ penalizes long service when information is already stale, while the quadratic term $\tfrac{1}{2}N^2$ captures the intrinsic AoI curvature associated with deterministic service. The third term enforces the semantic fidelity constraint through the virtual queue. Unlike the delay-optimal policy, decisions depend on instantaneous age $\Delta_k$ rather than backlog $Q_k$, leading to smaller $N$ when information is stale. Since AoI resets only upon semantic success, fidelity and freshness are tightly coupled. When semantic debt grows, the virtual queue $Z_k$ shifts the policy in \eqref{eq:aoi_policy} toward larger $N$ to reduce $\hat p_e(N)$ while maintaining $\bar{E}\le\varepsilon$.

\subsection{Selection of Minimum Feasible AoI}

For each arrival rate $\lambda$, we sweep the control weight $V$ to obtain the AoI-semantic fidelity tradeoff and select
\begin{equation}
\bar{\Delta}^\star(\lambda,\varepsilon)
=
\min_{V\ge 0}
\left\{
\bar{\Delta}(V)
:\ \bar{E}(V) \le \varepsilon
\right\},
\end{equation}
the minimum AoI over operating points satisfying $\bar{E}\le\varepsilon$.

This procedure parallels the delay analysis but targets freshness rather than backlog. The AoI-aware semantic controller provides a complementary operating regime to the queue-aware design of Section~\ref{sec:queue_sem_rate}. While the earlier policy is backlog-driven and prioritizes congestion mitigation, the present mechanism is freshness-driven and prioritizes timely semantic updates through the latent-dimension adaptation.

\subsection{AoI Performance Under Semantic Fidelity Constraints}

\begin{figure}[t!]
  \centering
 \includegraphics[width=0.8\columnwidth]{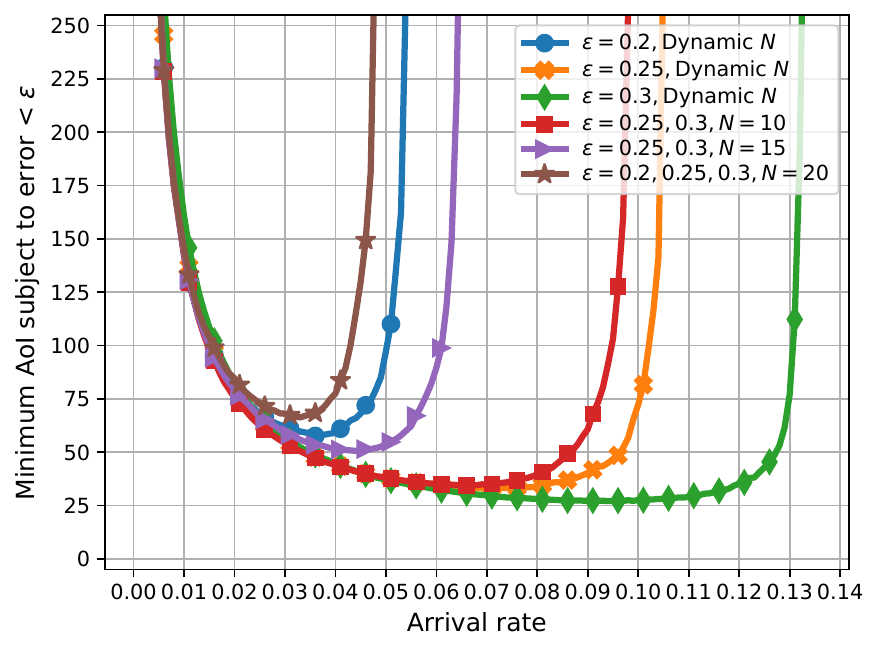}
  \caption{AoI vs. arrival rate.}
  \label{fig:aoi_vs_arrival}
  \vspace{-0.2cm}
\end{figure}

We next evaluate the AoI-aware semantic rate controller across varying traffic intensities. Fig.~\ref{fig:aoi_vs_arrival} shows the minimum achievable time-average AoI versus arrival rate $\lambda$ under error caps $\varepsilon\in\{0.2,0.25,0.3\}$ for both the dynamic latent-dimension policy and fixed-$N$ baselines. The results exhibit the characteristic U-shaped AoI versus load. At very small $\lambda$, the system is update-starved and AoI is dominated by inter-arrival time; as $\lambda$ increases, AoI decreases with more frequent updates. Beyond a critical load, service becomes the bottleneck and queueing dominates, causing AoI to rise sharply and marking the freshness stability boundary.

The semantic fidelity constraint strongly impacts the adaptive policy. Tightening the cap from $\varepsilon=0.3$ to $0.25$ and $0.2$ shifts the AoI curves upward and leftward because stricter fidelity forces more frequent selection of larger $N$, increasing service time and degrading freshness. The figure thus quantifies the intrinsic cost of fidelity in AoI-driven SemCom.

The dynamic latent-dimension policy consistently achieves the lowest AoI over the feasible region. At moderate load, it opportunistically uses smaller $N$ when the virtual fidelity queue is small, reducing service time while maintaining $\bar{E}\le\varepsilon$. Near heavy traffic, it increases $N$ only as needed to prevent fidelity violations, operating near the optimal freshness-fidelity balance. Fixed-$N$ baselines show clear feasibility limits. With $N=10$, only $\varepsilon\in \{0.25,0.3\}$ is feasible; $N=15$ meets $\varepsilon=0.25$ and $0.3$ over a wider range but struggles at $\varepsilon=0.2$; only $N=20$ satisfies all targets $\{0.2,0.25,0.3\}$, at the cost of substantially higher AoI. Thus, no fixed semantic rate achieves both low AoI and strict fidelity. The adaptive policy enlarges the AoI-stable region, as fixed-rate curves diverge earlier, while the dynamic controller maintains finite AoI over a broader $\lambda$ range by reducing the rate as age grows. Overall, AoI-aware semantic rate control enables lower AoI while automatically navigating the fidelity-freshness tradeoff.
\section{Conclusion} \label{sec:conclusion}
We considered semantic image communication over Rayleigh fading channels with an end-to-end learned encoder-decoder and identified the latent dimension as the key semantic-rate control variable coupling fidelity and service time. We showed that accuracy improves with latent dimension but saturates, while channel use (and latency) grows linearly, making fixed-rate designs inefficient. To address this, we developed semantic-fidelity-constrained rate controllers that adapt the latent dimension per update. The queue-aware policy minimizes delay under an average semantic error cap via a virtual fidelity queue and drift-plus-penalty selection, outperforming fixed-rate baselines and expanding the feasible load region. The AoI-aware policy targets time-average AoI under the same constraint and improves freshness across arrival rates. Overall, adaptive semantic-rate control provides a unified framework for balancing latency, fidelity, and freshness under dynamic traffic and channel conditions.
\bibliographystyle{IEEEtran}
\bibliography{references}
\end{document}